\documentclass[10pt,conference]{IEEEtran}
\usepackage{graphicx}
\usepackage{mathtools}

\usepackage{epsfig}
\usepackage{amssymb}
\usepackage{amsmath}
\usepackage{amsfonts}

\usepackage{algorithm}
\usepackage{algorithmic}

\usepackage{caption} 
\usepackage{subfigure}
\usepackage{fixltx2e}
\usepackage[font=small,labelfont=bf]{caption}
\usepackage{balance}

\usepackage{color}

\ifCLASSINFOpdf
\else
\fi
\begin{document}
%
\title{A Comprehensive Study on Load Balancers for VNF chains Horizontal Scaling}
%
%
%


\author{\IEEEauthorblockN{Jiefei Ma}
\IEEEauthorblockA{Imperial College, UK}
\and
        \IEEEauthorblockN{Windhya Rankothge}
\IEEEauthorblockA{SIIT, Sri Lanka}
\and
        \IEEEauthorblockN{Christian Makaya}
\IEEEauthorblockA{IBM, US}
\and
        \IEEEauthorblockN{Mariceli Morales}
\IEEEauthorblockA{UPF, Spain}
\and
       \IEEEauthorblockN{ Franck Le}
\IEEEauthorblockA{IBM, US}
\and
     \IEEEauthorblockN{Jorge Lobo}
\IEEEauthorblockA{ICREA \& UPF, Spain}
}
\maketitle

\begin{abstract}
%
We present an architectural design and a reference implementation for horizontal scaling of virtual network function chains. Our solution does not require any changes to network functions and is able to handle stateful network functions for which states may depend on both directions of the traffic. We use connection-aware traffic load balancers based on hashing function to maintain mappings between connections and the dynamically changing network function chains. Our references implementation uses OpenFlow switches to route traffic to the assigned network function instances according to the load balancer decisions. We conducted extensive simulations to test the feasibility of the architecture and evaluate the performance of our implementation.
%
%
\end{abstract}


%

\section{Introduction}
\label{introduction}

In this paper we propose a generic solution and architecture to horizontally scale virtual network functions (VNFs). 
The architecture relies on two load balancers: a {\it{master}} load balancer to handle traffic from the source to the destination of a connection, and a {\it{slave}} load balancer to handle the traffic in the reverse direction. 
These two load balancers rely on hashing functions to map connections to network function (NF) chains, and  ensure that both outgoing and incoming packets of each connection traverse the same sequence of network function instances.
In addition, we introduce algorithms for the load balancers to simultaneously add and/or delete network function chains, and evenly distribute the traffic among the network function chains especially considering that connections may be of different sizes and change characteristics over time.

We built a prototype to demonstrate the feasibility of the solution. We used a mix of virtual and physical components to run our experiments. To forward the traffic to the assigned network function instances, the load balancers add an identifier to each packet. The identifier is used by the switches for packet forwarding. In our implementation, we rely on OpenFlow switches for the data plane and use VLAN tags as identifiers. The number of tags needed for each additional network function chain is two. Hence, in a situation where, let's say, there are 3 instances of a network function for load balancing purposes, the load balancer use 6 tags. Our prototype relies on the open source PF\_RING network socket \cite{deri2004improvingPFRing} to capture incoming packets and add the VLAN tags. The throughput of the cards under this configuration was around 465~Mbps while the load balancer throughput was about 366~Mbps.  The results show that it takes less than 3 seconds for the traffic to be rebalanced when adding or removing a NF function instance. The experiments were run using standard installations of  Snort and Bro intrusion detection systems \cite{gerg2004managingSNORT,Bro}, but we will report only Snort experiments.  

The rest of the paper is organized as follows. The next section describes an operational scenario where the load balancers can be deployed and provides the general process flow that such deployment will go through. Section \ref{loadbalancers} describes how the load balancers handle the scaling of the network functions. Section \ref{referenceimplementation} describes our testbed and reference implementation. The evaluation and experimental results of using our prototype are presented in Section \ref{experiments}. Related work is discussed in Section \ref{RelatedWork}. Some general remarks and conclusions are presented in Section \ref{discussion}. 
%
%

\section{Operational Scenario}
\label{operational-scenario}

\begin{figure}
	\centering
	\includegraphics[width=2.3in]{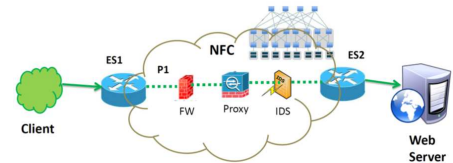}
	\caption{Original deployment of NF chain $P1$}
	\label{LBS1}
\end{figure}

\begin{figure}
	\centering
	\includegraphics[width=2.3in]{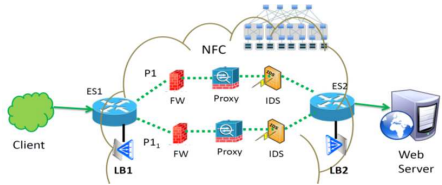}
	\caption{Load balancers deployment}
	\label{LBS3}
\end{figure}

To illustrate the problem, we consider the following scenario (Fig.~\ref{LBS1}): the network administrator of a web service wants the web traffic to be processed by a network function chain hosted in a Network Function Cloud Center (NFC). The chain consists of a firewall, a proxy, and an intrusion detection system (IDS).
Through the Management System (MS) of the NFC, resources are allocated and an initial instance $P1$ of the network chain is deployed. Web server requests are sent through the NFC, entering at the edge switch $ES1$, traversing  the VNF chain inside the NFC, and going out through the edge switch $ES2$, before reaching the Web server. The replying traffic from the Web server enters the NFC through the edge switch $ES2$, traverses the VNF chain inside the NFC in reverse order, goes out of the NFC through the edge switch $ES1$, and reaches the client. 
Horizontal scaling will replicate the chain as illustrated in Fig.~\ref{LBS3} with two load balancers, $LB1$ and $LB2$ to distribute the traffic going to the server and returning to the client. 


%
%

\begin{figure}
	\centering
	\includegraphics[width=2.3in]{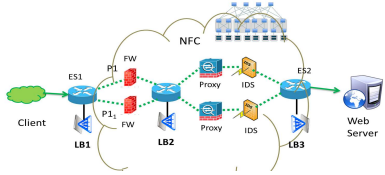}
	\caption{Load Balancers Chains}
	\label{LBS4}
\end{figure}

The load balancers are new network functions dedicated to the management functionalities of the NFC and they do not need to be deployed at the edges. The MS deploys and provisions pairs of load balancers such that each load balancer in the pair works in tandem with a switch (routing tables in the switch will be set according to the configuration in the load balancer) and in coordination with the other load balancer to balance the load and handle the traffic affinity in the presence of dynamic changes to the number of NF chains serving the traffic. In fact, load balancers can be deployed surrounding only the portion of the chain that needs to scale. For example, we can deploy $LB1$ after the firewall and before the proxy and $LB2$ after the proxy and before the IDS if only the proxy service needs scaling. It is also possible to deploy chained pairs of load balancers if there are differences in the scaling needs of the different NFs resulting in deployments as the one depicted in Fig.~\ref{LBS4}. In this figure, $LB1$ and $LB2$ coordinate the load balancing and traffic affinity of rescaling the firewall(s), and $LB2$ and $LB3$ coordinate the rescaling the proxy-IDS sub-chains. $LB2$ acts as a slave of $LB1$. $LB1$ decides how to split sessions between the firewalls and inform $LB2$ so that $LB2$ can split the returning traffic respecting session affinity. Similarly, $LB2$ is the master of $LB3$ and decides about session balancing among the proxy-IDS chains.

\subsection{Process flow}
%
We will describe the process flow of our implementation. Other options will be discussed in Section \ref{discussion}. 

The management system (MS) of the NFC upon receiving the service request for a new NF chain (see, for example, \cite{SIMPLE,rankothge2017optimizing, IEEESDN} for how this can be achieved), it will allocate and deploy resources as indicated in Figure~\ref{LBS1} with the caveat that of adding to the chain the master/slave pair of load balancers as depicted in Figure~\ref{LBS3}. 
The routing in $ES1$ will be set such as all the traffic coming from the client to be passed through the chain, $P1$ as well as the traffic coming back out from the chain going back to the client will be bridged to $LB1$. Similar routing settings are done in $ES2$ for $LB2$.

Hence, the MS knows how packets to be sent through $P1$ are identified by $ES1$ and $ES2$. This must be known by the load balancers as well so that they can take any packet received from $ES1$ or $ES2$ and modify it with the appropriate identification before passing the packet back to the switches. Therefore, the MS makes an asynchronous call to the master with the pair of IDs, one for the master to be used for the traffic that was passed to $LB1$ by $ES1$ coming from the client that will be passed back to $ES1$ to be routed to the proxy $P1$, one for the slave to do the equivalent work with the traffic coming through $ES2$ from the server to be sent to the IDS. 
The master synchronously calls the slave to pass the IDs and makes sure that both know about the IDs. 
From this moment on, all packets going from $LB1$ and $LB2$ to $ES1$ and $ES2$ are modified for the switches to route appropriately: Client - $ES1$ - $LB1$ - $ES1$ - $\cdots P1 \cdots$ - $ES2$ - $LB2$ - $ES2$ - Server.

The MS will monitor and analyze the traffic in the NFC and decide when the NF chain must scale, and as with initial request, it will allocate and deploy resources for the new child chain  as depicted in Figure~\ref{LBS3}. When the deployment is finished (the child path yet to be used), the MS will asynchronously call the master with the new pair of IDs, the master will call synchronously the slave to pass the new IDs and they will decide according to their owned load balancing policy how to modify the packets before passing them to the switches. 
The MS can asynchronously request statistics from the master and the master will collect data from the slave and pass the results to the MS about the distribution of traffic among the chains. The MS can also request the master to re-balance the load. The re-balance most have some coordination between the master and the salve. The master will coordinate the process.

\newcommand{\acts}{{\tt active\_Sessions}}

To maintain sessions affinity during dynamic re-balancing and changes of NF chains the load balancers maintain a table of \acts. We assume that a session or flow can be identified and mapped into a session {\tt key} regardless of the direction of the flow. This could be application dependent. For example, we can use the tuple (source IP, source port, destination IP, destination port) to identify a session in the outgoing flow and the tuple (destination IP, destination port, source IP, source port) to identify the same session in the returning flow, or more sophisticated techniques such as the one described in \cite{SIMPLE} to discover correlations between ingress and egress traffic of NFs. Regardless of the method, we need a network adapter in each load balancer that when it receives a packet {\tt p}, extracts its session {\tt p.key} and passes it to the load balancer. The load balancer returns a chain ID and the network adapter modifies the packet accordingly before putting the packet back into the network. To respect sessions affinity, the load balancers need to implement a consistent session function {\tt map()} that maps any packet {\tt p} of a give session to the same ID. Hence, in addition to having active sessions, the \acts~table also stores the IDs assigned to the sessions. Algorithm~\ref{LBAlgo} describes a general schema of how this computation can be done. The function {\tt get\_newID} will implement the load balancing policy. 

\begin{algorithm} \small
	\caption{Session-aware {\tt map(p:packet)}}
	\label{LBAlgo}
	\begin{algorithmic}
		\IF{{\tt p.key} is an active session}
		\STATE {\tt Update\_session(\acts[p.key])}
		\STATE RETURN \acts{\tt [p.key].id}	
		\ELSE
		\STATE {\tt instanceID = get\_newID(p.key) }
		\STATE {\tt Activate\_session(\acts , p.key, instanceId)}
		\STATE RETURN instanceID
		\ENDIF
	\end{algorithmic}
\end{algorithm}

The decision of removing a chain instance is also made by the MS. Similar to adding a new path, the MS will call asynchronously the master with the pair of IDs assigned to the path the MS wants to remove. The master synchronously will call the slave to communicate the IDs and a cool-down process will start in both the master and the slave. They will stop sending new sessions to the path being removed but they should keep using the path for active sessions already assigned to the path. The MS polls the master and the slave  and ask whether the path is active (i.e., if there are still active sessions assigned to the path). When both, the master and the slave, respond that the path is not active the MS can re-use the resources of the chain for other tasks. The master and slave must access the \acts~table to decide if a path is still active avoiding race conditions affecting updates to this table. Thus, the implementation of {\tt map} must ensure that reads are not done when changes are being made to the table - a direct solution is to set {\tt map} as a critical region that does not allow concurrent reads to the table. Note that the re-balancing that will affect the behavior of {\tt get\_newID} when adding or removing instances, or at any time that the MS requests it, can be done concurrently since it does not read the \acts~table. Session management overrides balancing. The challenge now is how to make sure that IDs will be assigned consistently in both the master and the slave. We will discuss this in the following section.

\section{Load Balancers}
\label{loadbalancers}

The basic functionalities of a master-slave pair of load balancers are: (1) given a session key $k$ to consistently return an ID so that the packet is sent through the right path, (2) trying to distribute the traffic evenly among the chains, and (3) to simultaneously deal with additions and deletions of available chain IDs.   

For (1) and (2), we have designed the following  strategy. We have identical vectors of $L$ buckets, with $L$ much larger than the possible number of IDs, in each load balancer. The load balancers assign IDs to buckets in proportion to how much traffic they like to see going through each chain. In the initial case, when there is no information, if there are $N$ different IDs, the load balancers assign the same ID to $L/N$ buckets; the same order of assignment is done in both load balancers. The buckets are shuffled in the vector using a pseudo-random generator starting with a given seed $d$ known to both load balancers, and any shuffling  must be done in both balancers, the order is irrelevant, it can happen in parallel but and no processing of traffic before both balancers have done it. We also have a function $h$ used by both load balancers that maps session keys to positive integers. When a load balancer receives a packet with session key $k$ the packet is sent to the chain with the ID stored in the bucket at position $h(k)\; mod\; L$ of the vector. In our implementation, we use a hash function to get the integer. If sessions and session lengths are uniformly distributed and the traffic flow does not fluctuate over time this fully addresses (2). To deal with variation of session lengths as well as (3), scaling (i.e., traffic fluctuations) we have implemented the following strategy. 

For a given time-window $\tau$, let $t_i$ be the total number of incoming and outgoing bytes going through chain $n_i$. Thus, the total traffic in the time-window is:
$
T = \sum_{i=1}^N t_i.
$
Let $l_i$ be the number of buckets assigned to the corresponding chain $n_i$. Hence, 
$
L = \sum_{i=1}^N l_i.
$
We can approximate the probability $p_i$ of the hash function assigning a (random) session to chain $n_i$ with:
$
p_i = l_i / L.
$
If all sessions are uniformly distributed and have the same length (e.g., in a controlled environment), we would have:
$
T p_i = t_i.
$
However, since some sessions may have many more packets than others or packets of different sizes, we assume the existence of a session bias $b_i$ for each chain $n_i$, such that,
$
T b_i p_i = t_i.
$
Using $b_i= t_i / (T p_i)$, we derive equations to re-allocate traffic loads as follows. First, to simply redistribute the traffic load whenever the traffic changes, we readjust to new probabilities $p_i^{new}$,  assuming that the near future will be similar to the past using the equation:
$
T b_i p_i^{new} = T / N
$,
which written in terms of our inputs we get:
\begin{equation*}
p_i^{new} = \frac{\frac{T p_i}{N t_i}}{\frac{T}{N}\sum_{j=1}^N \frac{p_j}{t_j}} 
\end{equation*}
The denominator is to normalize the proportions to 1 since we want: 
\begin{equation*}
\sum_{j=1}^N p_j^{new} = \frac{T}{N}\sum_{j=1}^N \frac{p_j}{t_j} = 1 
\end{equation*}
Hence,
\begin{equation*}
p_i^{new} = \frac{p_i}{t_i\sum_{j=1}^N \frac{p_j}{t_j}} 
\end{equation*}
Following the same approach of recalculating the probabilities, if a new chain $n_{N+1}$ is added, we adjust the probabilities for all $n_i$, $i\leq N$ by:
\begin{equation}\label{nplusone}
p_i^{new} = \frac{N}{N+1}\frac{p_i}{t_i\sum_{j=1}^N \frac{p_j}{t_j}} 
\end{equation}
%
and set $p_{N+1}^{new} = 1/(N+1)$. If, on the other hand, the chain $n_k$ is removed, for all $k\neq i$, we set:
\begin{equation*}
p_i^{new} = \frac{p_i}{t_i\sum_{j\neq k} \frac{p_j}{t_j}} 
\end{equation*}
and set $p_k^{new} = 0$. With these new probabilities we recompute each $l_i$ to be $p_i^{new} L$, and re-generate the bucket vector assignment. Note that the unbalanced distribution of traffic from one time window to the next is proportional to the difference between the biases from one time window to the next, i.e., $b_i^{\tau +1}/b_i^{\tau}$. With a good hashing function, IDs should be distributed uniformly through the sessions but unbalancing can occur if the amount of traffic among the sessions is no uniformly distributed. Additionally, if sessions with the same IDs are not distributed uniformly over time sessions with the same ID will use the same path.  One needs to be careful about deciding the size of the window. In a big window the distribution of traffic might average out over the time of the window but there might be portions of the window in which the traffic is skewed in one direction that is ``compensated" by another portion of the window where the traffic is skewed in another direction. One can use windows of small sizes since this will limit possible changes in biases but this might burden the operation of the load balancer if it expends too much time doing re-balancing operations and coordinating re-shuffles. The right window size could be decided by a hysteresis analysis of the traffic but it is outside the scope of this paper.  

For maintaining the affinity of session, each load balancer has a table of sessions storing the session {\em status} $s$ that is defined as the last time the load balancer saw a packet from that session. It also stores the last chain ID, $i$, assigned to that session. Given a fixed session timeout $\rho$ available to both load balancers, a load balancer considers the session active if $s + \rho$ is lager that the current time. If a packet arrives when a session is active the session status is updated and the packet is sent to the current assigned chain $i$. If the session is not active, the status is updated and the packet is sent to the chain in position $h[k]\; mod\; L$ of the bucket vector and $i$ is updated. Hence, Algorithm~\ref{LBAlgo} is instantiated in our implementation Algorithm~\ref{LBAlgo1}.
\begin{algorithm}\small
	\caption{Session-aware {\tt hashing(p:packet)}}
	\label{LBAlgo1}
	\begin{algorithmic}[1]
		\IF{{\tt p.ses} is an active session}
		\STATE $sesStatus \leftarrow activeSessions[key(p)]$
		\STATE $sesStatus.lastTimestamp \leftarrow p.timestamp$
		\STATE RETURN sesStatus.tag	
		\ELSE
		\STATE $InstanceID \leftarrow buckets[ hash(key(p)) \% buckets.size ]$
		\STATE $newSesStatus.lastTimestamp \leftarrow p.timestamp$
		\STATE $newSesStauts.tag \leftarrow InstanceID$
		\STATE $activeSessions[key(p)] \leftarrow newSesStatus$
		\STATE RETURN InstanceID
		\ENDIF
	\end{algorithmic}
\end{algorithm}

Because the load balancers share the seed for the pseudo-random generator the coordination is minimal.  We are following the model in \cite{SIMPLE} where traffic can be dynamically redirected to the appropriate NF in the network and the ideas from \cite{IM15,rankothge2017optimizing} where a NFC management system is able to deploy new VNF chains under different optimization conditions.  This management system also picks the IDs of new chains and in our specific case it can also communicate the IDs to the master load balancer. 

It is theoretically possible that active sessions are not synchronized because of delays between the arrival of packets to the load balancers, this can be quickly corrected by the master if it observes a packet returning with an ID different from the ones it locally has for packets of the same session.

\section{Reference Implementation}
\label{referenceimplementation}

\begin{figure}[h]
	\centering
	\includegraphics[width=2.5in]{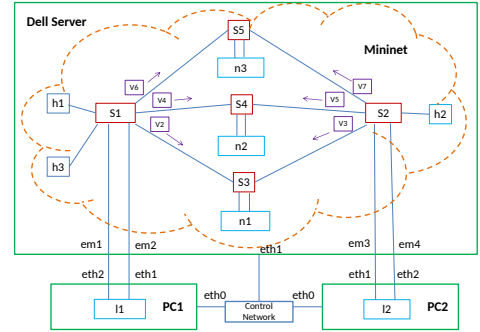}
	\caption{Testbed Architecture}
	\label{testbed}
\end{figure}

To test our proposal we needed a reference implementation of a full operational testbed. We had to decide how to implement two components of the architecture. We needed a: (1) network where we could reroute traffic dynamically and (2) a concrete implementation of a network adapter. For (1), we have used OpenFlow \cite{openflow}. For (2), the load balancer will take a session to be all the packets with the same set of pairs (source IP:port, destination IP:port) in the headers that pass through the load balancer for which any two consecutive packets with the same set are not separated by more than a fixed time window $\tau$. The architecture of the deployment is depicted in Fig.~\ref{testbed}.  
We have three physical machines: a PowerEdge R430 Dell Server with 12 cores, 32~Gb of memory and four 1~Gb Ethernet cards, and two ALDA+ PCs both having a dual core 6320 Intel CPU and 4~Gb of memory. The PCs have three network cards two of which were able to support the open source PF\_RING network sockets \cite{deri2004improvingPFRing}. This was used to implement the load balancer network adapter. Standard sockets are reasonable to monitor traffic but too slow when doing packet manipulations. Other alternatives exist such as DPDK, but we decided to use PF\_RING based on the available hardware in our testbed environment. The network adapter for the load balancers was built using PcapPlusPlus \cite{PcapPlusPlus}, a lightweight C++ packets manipulator library, which can be used to parse and create or manipulate packets and is compatible with with PF\_RING.  

We describe next the rest of the testbed components before explaining how OpenFlow is used. In Fig.~\ref{testbed}, the cloud inside the Dell server represents a Mininet deployment. It has 5 switches and 6 servers. The servers named $n_i$ will run network functions. All the switches, the $n_i$ servers and PC1 and PC2 are assumed to be under the NFC management. The $h_i$ servers are assumed to be outside the NFC and will be sending and receiving traffic. Two ports of Switches $s_1$ and $s_2$ are directly associated to the physical cards of the Dell server to send and receive traffic from the load balancers running on the PCs. The third network cards in PC1 and PC2 and an USB network interface in the Dell server (eth1) are used for the control traffic of the MS. The load balancer $l_1$ running in PC1 will act as a master of the  slave load balancer $l_2$ running in PC2. Server $h_2$ will be running a web server and clients will connect from $h_1$ and $h_3$. 

\subsection{OpenFlow Management}
The NFC management uses a unique pair of VLAN tags for each NF chain and it will install OpenFlow rules in the appropriate switches to route the traffic inside the NFC. The idea of using VLAN tags for routing is not new and has been successfully used in \cite{SIMPLE,IM15}. In our testbed, we will use three chains with a single NF in it: one chain will use the pair of VLAN tags $(2,3)$, the second chain will use the pair $(4,5)$ and the last chain will use the pair $(5,6)$. 

When a master is brought up, it is informed of the address of its slave through the control plane channel and does a handshake with the slave, which means that the slave needs to be brought up first. At any moment the MS can send pairs of IDs (VLAN tags in our case) to the master and it will pass the information to the slave and they will automatically start using the new IDs for balancing. Similarly, pair of IDs can be removed at any moment through the master. We have implemented a polling mechanism to ask the load balancers what IDs are active since old sessions can linger for some time after removing a pair of tags. The load balancer network adapters take traffic coming from interface eth2, extract the (source IP:port, destination IP:port) pair and gets a ID from the load balancer for this tuple. The network adapter takes that ID and adds it to the packet as a VLAN tag and puts the packet in the network card in eth1. 

The load balancers assume that all the appropriate forwarding rules associated with the IDs they are aware of have already been installed in the switches. For example, assuming that port numbers are assigned counter-clock wise starting from the port connected to $h1$, switch $s_1$ has flow rules that say: {\it{any packet coming into port 1 or 2 will be sent out from port 3, any packet coming into port 4 will be sent out from port 5 if the VLAN tag is 2, to port 6 if the VLAN tag is 4 and to port 7 is the VLAN tag is 6, and finally, any packet coming with VLAN tags 3, 5 or 7 is sent out from port 3}}. If more switches were in the path from $s_1$ to switches $s_3$, $s_4$ or $s_5$, similar rules would be needed.  Switches $s_3$, $s_4$ and $s_5$ will drop the VLAN tags before passing the traffic to the network functions and will add the same tags back to the packets coming out from the network function. For example, a packet coming into port 1 in $s_3$ with VLAN tag 6 will be sent out from port 2 without the tag, all the packets coming into port 3 will be sent out from port 4 after adding the tag 6. Packets coming into port 4 with tag 7 will be sent out from port 3 without tag, and any packet coming into port 2 will be sent out from port 1 with tag 7. This way, the network function will see the packets without any modification, even if the VLAN tags were originally in the packet. This is possible by the stack mechanism that OpenFlow implements to handle VLAN tags.

\section{Experiments}
\label{experiments}

We have conducted two sets of simulations. One to test the performance of the load balancers independent of the network functions, and a second set to test the performance of the load balancers with traffic that was slowdown by the limited capacities of network functions. 

\subsection{Load Balancer Performance}

The first test that we did was a clean throughput test over the testbed. All the Mininet links were set to have bandwidths of 1~Gbps and we set Linux bridges between the interfaces of $n_1$ as well as between network interface cards eth1 and eth2 in both PC1 and PC2. We also set OpenFlow rules in $s_1$, $s_2$ and $s_3$ so that any traffic between $h_1$ or $h_3$ and $h_2$ will follow the path ($s_1$, PC1, $s_1$, $s_3$, $n_1$, $s_3$, $s_2$, PC2, $s_2$, $h_2$). 
Using iperf and the proper parameters in httperf, we consistently got throughput over 900~Mbps. Next, we replace the Linux bridges in PC1 and PC2 with bridges using the vanilla version of PF\_RING and PcapPluplus to parse the packet arriving into eth2 and putting it back into eth1. In this case the average throughput went significantly down to an average of 465~Mbps. Better configurations that could have be done to improve the performance of PF\_RING with considerable more coding (e.g., not using PcapPlusPlus) but we found this throughput adequate for our goals.

\begin{figure}
	\centering
    \includegraphics[width=2.6in]{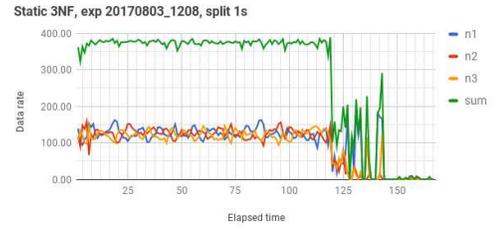}
	\caption{Throughput with static number of NFs}
	\label{static}
\end{figure}

\paragraph{Static scenario} Having established the throughput of the testbed, we next tested the performance of the load balancers in static situations. We ran three sets of experiments fixing the number of servers where network functions will be running. We ran 5 experiments using httperf generated traffic consisting of 8000 http connections at a rate of 75 connections per second, each connection fetching a 0.75 MB file.  For this initial case, we just created a Linux bridge in each of the $n_i$ servers (i.e., void network functions that just passed the traffic through) to only measure the performance of the load balancers. The same five runs where done simulating one, two or three (void) network functions for a total of 15 runs. The traffic was captured in each $n_i$ using tcpdump and splitting the pcap files into files covering intervals of 1 second each. Figure \ref{static} shows one of the plots for one of the five runs in the case of three network functions. All plots were similar to these ones in shape. We computed the average throughput of the five runs from second 20 to second 110 to eliminate the initial and final jitters of the data. The average throughput with a single network function was 367~Mbps. The throughput with two network functions was 366~Mbps splitting the average throughput evenly between $n_1$ and $n_2$ (not statistically significant difference from having a single network function). The throughput with three network functions was slightly better with an average of 372~Mbps with little statistical differences, splitting the throughput into 125~Mbps for $n_1$ and $n_2$ and 122~Mbps for $n_3$. The distribution of traffic was in average 33\% per network server during the full run, even outside the [20, 110] time interval.  

\begin{figure}
	\centering
	\includegraphics[width=2.4in]{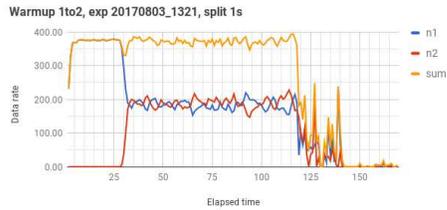}
	\caption{Throughput with dynamic increases of NFs}
	\label{warmup}
\end{figure}

\paragraph{Warm-up scenario} The next set of experiments tested the amount of time it took for the traffic to be balanced after a new network function has been added. The traffic generated was the same as in the static simulations: 8000 http connections sent at a rate of 75 connections per second fetching a 0.75 MB file. There were two sets of experiments, one set in which each experiment started with one network function running and the second set each experiment starting with two network functions running before starting to sent traffic. Again, these experiments used void network functions just letting traffic pass through.  Traffic was being monitored in interface eth2 of PC1 and 25 seconds after  the first packet with destination server $h_2$ was detected a call was made to the master with two new VLAN tags initiating the redistribution of the traffic with the extra network function server. Each set of experiments consisted of five runs for a total of 10 runs. Figure \ref{warmup} shows the plot of one of the five runs for  the case of two network functions. The plots for the other cases followed the same pattern. The plot shows the traffic throughput of each network function server at every second.

\begin{figure}
	\centering
	\includegraphics[width=2.4in]{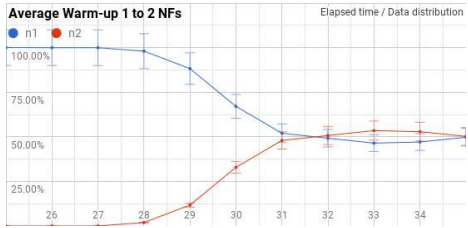}
	\caption{Average distribution of traffic among NFs}
	\label{warmuptime}
\end{figure}

We again took the average throughput in the [20, 110] time interval which includes the time when the tags were passed to the master load balancer. The throughput was similar to the static cases, 374~Mbps for the case of moving from 1 to 2 network functions and 363~Mbps for the case of 2 to 3 network functions. Figure~\ref{warmuptime} shows the ratios of the traffic distributions in average between the different network function server. We zoom-in into the plot to show the interesting time interval [25, 35] where all the re-balancing of the traffic happens.  

The precision of the measurement is in seconds. 
Hence, if tags are passed to the master after 25 seconds have elapsed, we will start seeing traffic changes after 1 sec. At the 25 sec. mark, half of the new connections will be directed to the new network function when we move from 1 to 2 network functions, and a third, when we move from 2 to 3 network functions. In the first couple of seconds less than 1\% of the traffic passes through the new network function, after that the plots start to show the changes. From the httperf reports of the static experiments, we gathered that the median duration for the connections was 6 seconds. Each http connection generated by httperf uses a different source port. Hence, a connection corresponds to a session and analytically, given that the policy we use when adding a new instance to a set of $N$ instances is to redirect in average $1/(N+1)$ of the traffic to that new instance (see Eq.(\ref{nplusone})), we can predict that after 6 seconds all sessions started before the 25 sec. mark should have finished, and the traffic should be balanced. 
Fig.~\ref{warmuptime} confirms the analysis. We show the 10\% error bars -- after 32 seconds error bars start to overlap. 

\begin{figure}[h]
	\centering
    \includegraphics[width=2.6in]{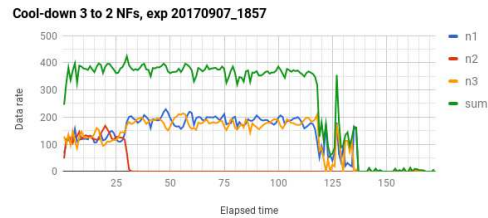}
	\caption{Throughput with dynamic decreases of NFs}
	\label{cooldown}
\end{figure}

\paragraph{Cool-down scenario} The experiments where network functions were removed were symmetric to experiments in the warm-up scenarios: Two set of experiments where we started with two or three network functions and after 25 seconds a request to the master load balancer was passed to remove one of the network function paths. No re-balance is needed when going from two to one network function but the plots show the cool-down time needed to have no traffic going through the deleted network function. Typical behavior of the traffic distribution is depicted in Figure \ref{cooldown}. The figure shows the cool-down from 3 to 2 NFs.

\begin{figure}
	\centering
    \includegraphics[width=2.4in]{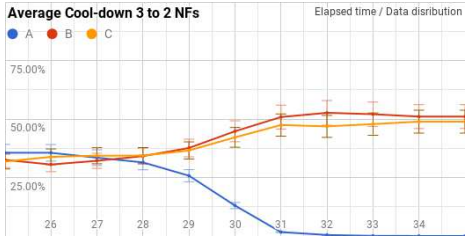}
	\caption{Average distribution of traffic among NFs}
	\label{cooldowntime}
\end{figure}

The throughput did not get affected. The average throughput for the [20, 110] time interval was 365~Mbps for the two-to-one experiment set and 368~Mbps for the three-to-two experiment set. Figure \ref{cooldowntime} shows that the average cool-down time is 6 seconds (the median length of the sessions) and that, for the three-to-two case (shown in the figure), traffic is distributed evenly between the two remaining network functions (in the experiments, we varied the network function removed to make sure there were no biases in the re-balancing) -- The 10\% error bars are constantly overlapping.

\subsection{Simulations with Snort}

We now shift gears in this section from stress test evaluation results to experiments with a real network function. For these experiments we ran the same set of experiments but this time running Snort in $n_1$, $n_2$ and $n_3$. We configured Snort in in-line mode with simple rules to detect potential SQL injections attacks, and
injected URLs that contained SQL statements as they would appear in some SQL injection attacks using curl calls. 
{Alerts were collected in Snort logs to verify that all the injections were detected, and that injection traffic was  evenly distributed among the NFs.}\footnote{We also ran simulations with Bro, but given its detection and not prevention nature, processing traffic with Bro did not affect network throughput.} 

\begin{figure}
	\centering
	\includegraphics[width=2.6in]{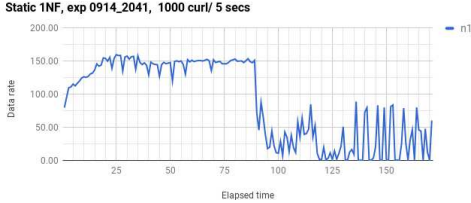}
	\caption{Snort throughput with 20K SQL injections}
	\label{staticcurl}
\end{figure}

We first measured the throughput of Snort by running experiments with the same settings as in the static experiments with a single network function and no curl calls. The average throughput went down to 162~Mbps. We repeated the experiments but this second time, we generated using httperf 3000 connections at a rate of 40 connections per second transferring a 1 MB file per connection with the same average results. We reduced the number of connections because we were getting to the limits of file descriptors needed to run the simulation in the Dell server with 8000 connections at the 75 connections/sec rate, and the number of file descriptors needed crossed the limit when add the curl calls to the httperf traffic. We ran experiments injecting 600 and 1000 URLs every 1, 3, and 5 seconds, reaching up to 80,000 URL injections in some of the runs. We got some but consistent decrease in throughput (from about 162 to 152 Mbps) when we had more than 5,000 URL injections. The plots looked more jagged than without URL injections -- see Fig.~\ref{staticcurl}. We suspect this is caused by the short-lived curl sessions. The intervals to calculate the average values were adjusted to be over [20,80] because the traffic started to decrease earlier than in the previous experiments. 

\begin{figure}
	\centering
	\includegraphics[width=2.6in]{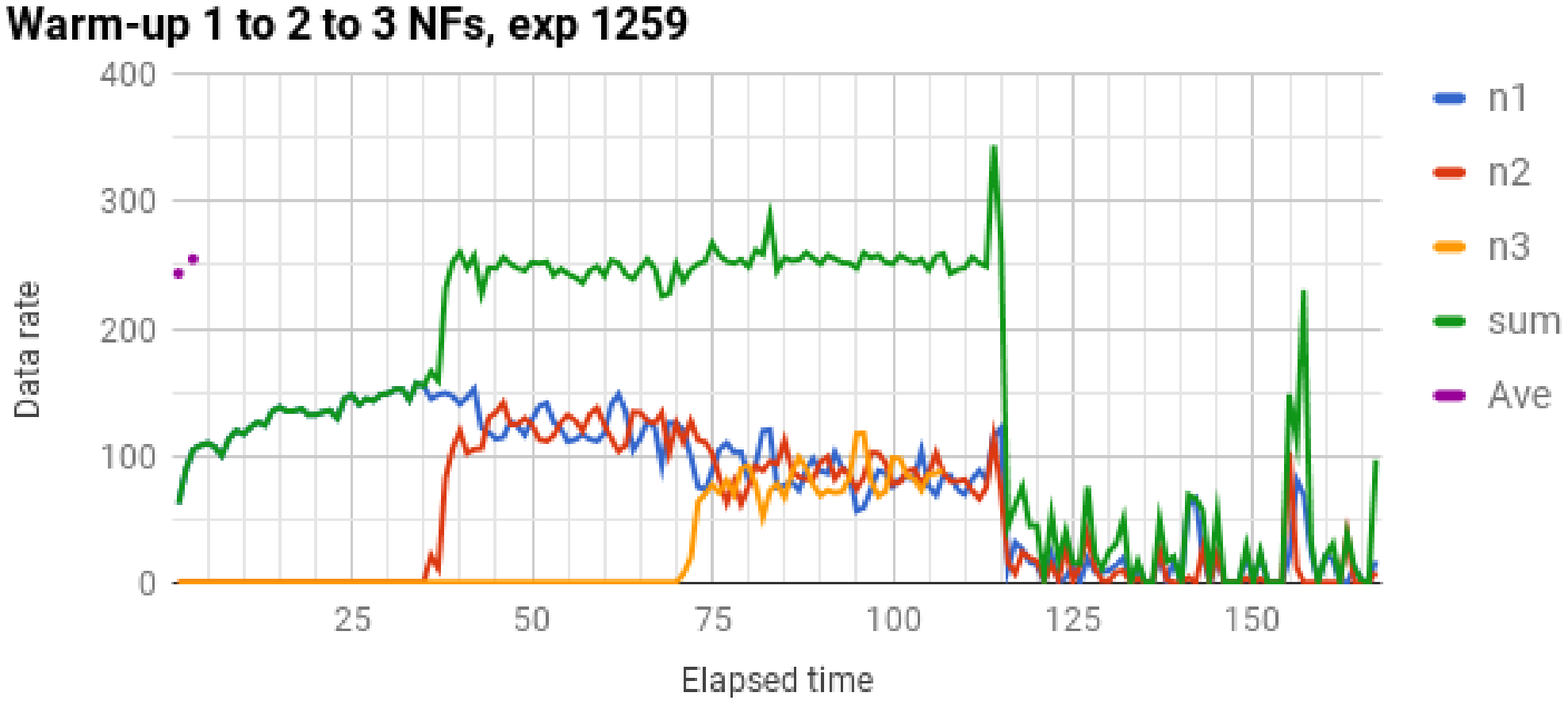}
    \includegraphics[width=2.4in]{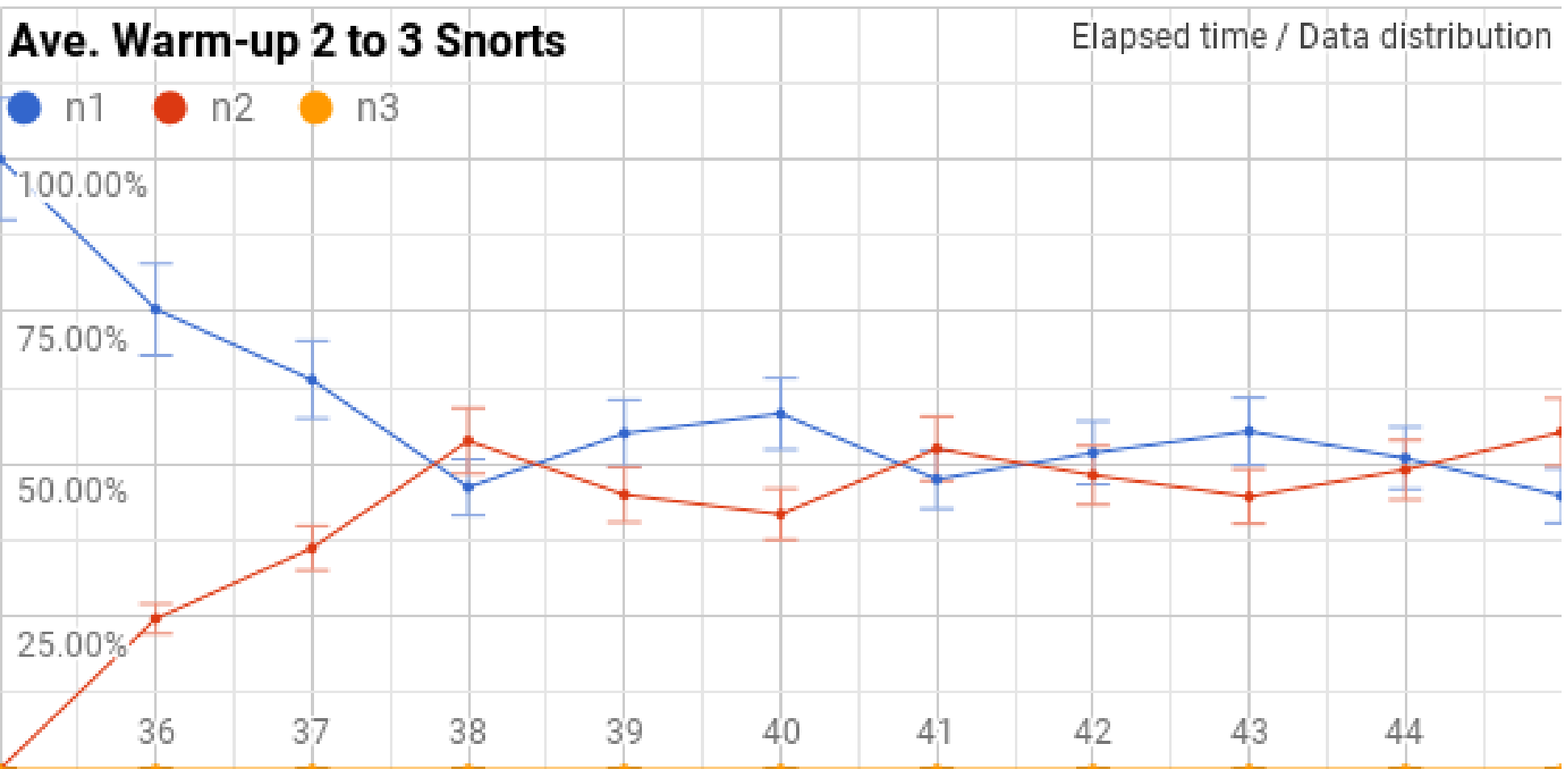}
	\caption{Dynamic increases of Snort instances}
	\label{warmupcurl}
\end{figure}

\paragraph{Warm-up} we ran the same series of warm-up experiments as in the case of void network functions. Additionally, we ran a set of five simulations in which we started with one instance of Snort running, then after 35 seconds had elapsed, we sent a pair of tags to the master load balancer to add a second Snort instance, and then, 35 seconds later, a second pair to add a third instance. The results for the combined experiments followed the same pattern than the individual warm-ups. Because of space limitations only plots for the combined experiments are shown. Since we needed a longer experiment to be able to add 2 pairs of tags in the same run we use the same httperf traffic as with the static experiment sets: 8000 http connections generated at a rate of 75 connections/second, each connection transferring a 0.75MB file from server to client. This traffic, generated using httperf from host $h_1$, was accompanied with an average of 10K curl calls generated from $h_3$ with SQL-like injection attacks. URL injections were limited to 10K to avoid exhausting file descriptors.  The graph in Fig.~\ref{warmupcurl} is typical for all the cases. 

The top graph shows the throughput over time for one of the five runs. We measured the throughput in the intervals [10, 35], [37, 70], and [75, 110] averaging 148, 247, and 277 Mbps respectively. The second graph plots the time it took, in average, for the traffic to re-balance after the master receives the request to bring up the path to the second instance of Snort after 35 seconds have elapsed. The graph zooms-in to the 35 second mark. It takes about 3 second for the traffic to be balanced - 10\% error bars are shown in the plot. The graph around 70 seconds is similar but not shown.

\begin{figure}
	\centering
	\includegraphics[width=2.6in]{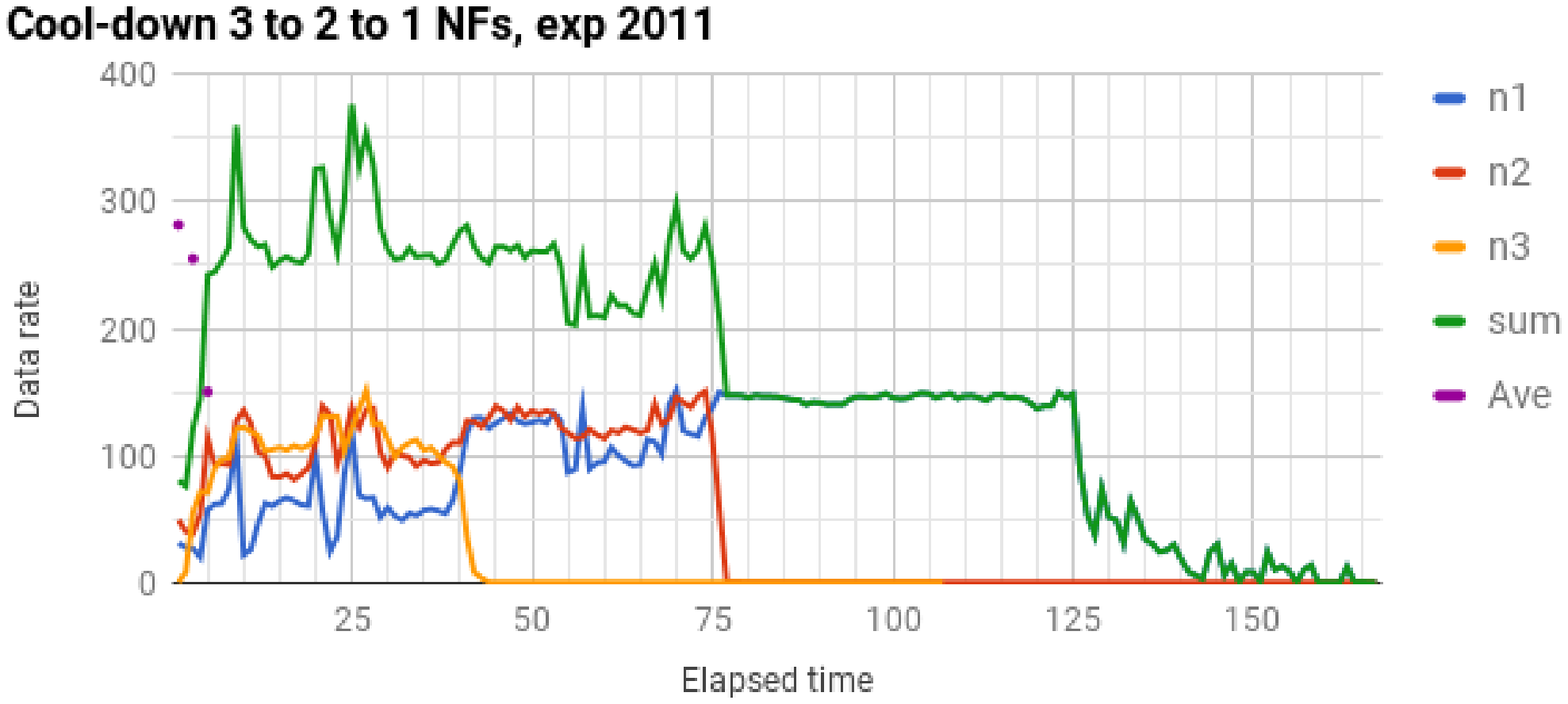}
    \includegraphics[width=2.4in]{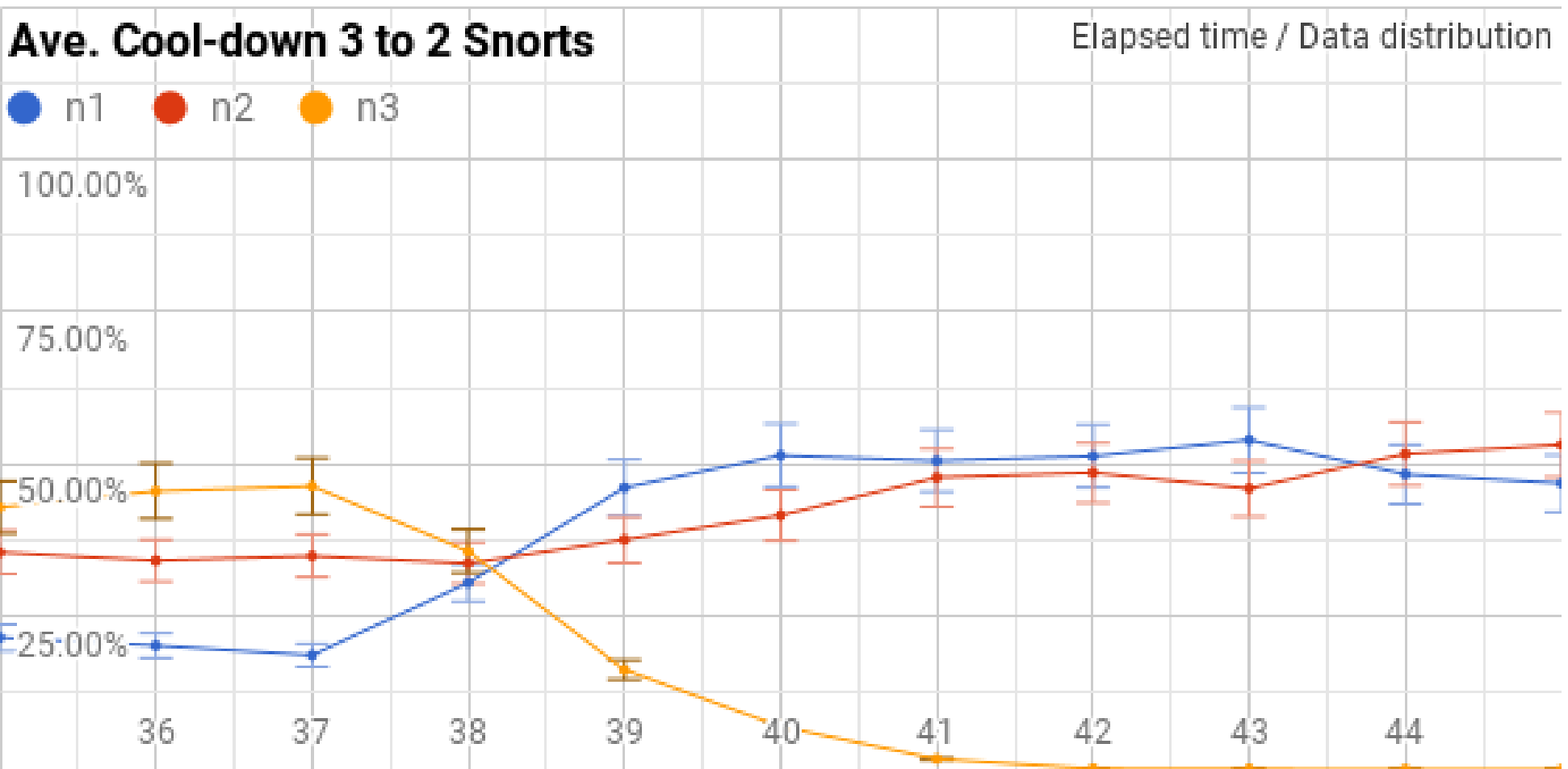}
	\caption{Dynamic decreases of Snort instances}
	\label{cooldowncurl}
\end{figure}

\paragraph{Cool-down} we ran a set of five cool-down simulations starting with 3 instances of Snort running. After 35 seconds had elapsed, we sent a request to the master to remove one of the instances, and 35 seconds later we sent another request to remove a second instance. The traffic load was as with the combined warm-up. Typical results are illustrated in Fig.~\ref{cooldowncurl}.
The average throughput in the intervals [10,35], [40,70], and [75,110] were 270, 244 and 152 Mbps showing a symmetric behavior from the warm-up scenario. The cool-down time was about four seconds - indicated by the moment when the 10\% error bars of the Snort instance stop overlapping with error bars of the reminding instances. 

\section{Related Work}
\label{RelatedWork} 

Perhaps the first realistic demonstration of the use of SDN for load balancing  NF traffic appeared in \cite{SIMPLE}. The paper discusses an implementation of a new controller called SIMPLE to balance the work load across a fixed set of NFs  -- no for scaling. To alleviate heavy network traffic and reduce the risk of a single  NF server being overloaded, SIMPLE  adopted dynamic load balancing mechanisms and distributed the traffic across multiple physical servers using dynamic updates of flow tables in the switches. Although scaling was not the goal, all the ingredients to develop VNF scaling management for the cloud were present. This was demonstrated in \cite{IM15}. The latter work does not discuss how to do load balancing but to optimally select networking and computer resources for NF. We will compare the load balancing approach of SIMPLE with ours in Section~\ref{discussion}.

Most of the existing scaling solutions that handle session affinity depend on state migration techniques: moving the relevant state from one instance to another. Frameworks like Split/Merge \cite{Split} and OpenNF \cite{gember2014opennf} facilitate fine-grained transfers of internal NF state to support fast and safe reallocation of flows across NF instances. In these frameworks, a scenario-specific control application decides: (1) when internal NF state should be moved; i.e., after a new NF instance is launched; (2) what subset of state should be moved; this is usually defined in terms of a flow space \textit{fspace}, i.e., all state pertaining to flows originating from a particular subnet; and (3) between which pair of NF instances the transfer should occur. A central controller then asks the source NF instance to export the state pertaining to flows in \textit{fspace}. This state is provided to and imported by the target NF instance. In Split/Merge, the state is transferred directly from source NF instance to the target NF instance, while in OpenNF the state passes through the controller. Finally, the controller updates the forwarding state in the SDN switch, such that traffic in \textit{fspace} is now forwarded to the target NF instance. The drawback of these two approaches is that NF code must be modified prior to support such export/import operations. 
Another limitation is that they require a large number of rule sets in the switches and it is not clear whether the methods as presented will scale.

E2 , described in \cite{palkar2015e2}, is a VNFs scheduling framework that supports traffic affinity based on NF placement and dynamic scaling while trying to minimize traffic across switches. It introduces a high-performance and flexible data plane where Click \cite{NFV3} modules (i.e., classifier and TCP re-constructor, etc.) are embedded to accelerate packet processing. Although modifications to the NFs are not strictly required, they should use the API exported by the data plane to achieve the full benefit of the framework. As us, E2 uses hashing functions applied to packet headers to partition and load balance the traffic but it is not able to re-balance in cases where traffic changes shape but not size. E2 also uses a novel migration avoidance strategy in which the hardware and a software switch act in concert to maintain the affinity.
However,  \cite{palkar2015e2} does not describe how bi-directional flow sessions are handled.

\section{Final Remarks}
\label{discussion}

We have presented a load balancing methodology for the management of horizontal scaling of NF chains that does not require changes to the NF code. We also developed a prototype reference implementations to illustrate the feasibility of the proposed solution and conducted extensive simulations to assess the performance. There were several implementation decisions that were taken to complete the reference implementation that can be varied. We briefly review a few of them.

Our process flow (see Sec.~\ref{operational-scenario}) introduces load balancers during the initial deployment of the service request. Under this approach, if we would like to split the chain at any point (see Fig.~\ref{LBS4}), one would need to provision load balancers between every NF. Alternatively, load balancers can be added as needed. This is possible by doing the scaling in two steps. First, we can start the load balancers and make the changes to the forwarding tables in the switches, and then after all is set, the traffic is processed by the load balancers as described in Sec.~\ref{operational-scenario}. This, of course, delays the first scale-out but it is possible if  scale-outs are predicted with sufficient time. 

Rajagopalan et al. \cite{Split} ran traffic simulations containing, among other traffic, a few long sessions. The point was to produce scenarios where few sessions remain active for a long time after the MS has decided to scale-in the chain through which these long sessions are being served. This implies that resources will not be released until the sessions finish. One may address this issue by using a mix of scaling methods. For example, one can use vertical scaling techniques to reduce the resources allocated to the chain to support the few remaining sessions, or do a migration of what it could be a small state.   

There is a possible configuration in which a single load balancer plays the role of master and slave for the instances of the same chain and avoid the synchronization: one could set a load balancer before the chain to allow scale in/out, and route both incoming and outgoing traffic through the same load balancer. This would increase the traffic passing through the switch in front of the load balancer by 50\%, but it would be an alternative to avoid the synchronization. This could be the method used in \cite{palkar2015e2}. This is also similar to the deployment presented in \cite{SIMPLE}, with the main difference being that the load balancing in \cite{SIMPLE} is done by an OpenFlow switch in front of every NF and the load balancing policies are implemented by the controller. Considering the load balancing as a management NF as we do, the dependency between the implementation of load balancers and the network data flow control is limited to the ability to add and remove information  to the packets to identify the data path and thus, many load balancing methods and sessions definitions adapted to the different NFs can be deployed.




\ifCLASSOPTIONcaptionsoff
  \newpage
\fi



\bibliographystyle{IEEEtran}
\bibliography{network}

\section*{Note}
All code will be made available under an Open Source license by the time of the publication. Access can be given to the reviewers if requested.
\end{document}